\documentclass[preprint]{revtex4-2}

\usepackage{graphicx}
\usepackage{newtxtext}
\usepackage{newtxmath}
\usepackage{natbib}
\usepackage{bm}
\usepackage{subcaption}
\usepackage{mathrsfs}
\usepackage{hyperref}
\usepackage{framed}
\hypersetup{
    colorlinks = true,
    urlcolor   = blue,
    citecolor  = black,
}

\newcommand{\RomanNumeralCaps}[1]
\linenumbers

\def\bw{{\bm \omega}}
\def\bk{{\bf k}}
\def\eps{\varepsilon}
\def\be{{\bm \zeta}}
\def\re{{\rm e}}
\def\ri{{\texttt i}}

\def\bth{{\bm\theta}}
\def\qand{\quad\mbox{and}\quad}

\def\Zh{\widehat{Z}}

\def\D{{\rm D}}

\begin{document}

\title{Modulation leading to frequency downshifting of water waves in the vicinity of the Benjamin-Feir transition}

\date{\today}

\author{Daniel J. Ratliff}
\affiliation{Department of Mathematics, Physics and Electrical Engineering,
Northumbria University, Newcastle upon Tyne, NE1 8ST, UK}
\author{Olga Trichtchenko}
\affiliation{Department of Physics and Astronomy, The University of Western Ontario, London, N6G2V4, Canada}
\author{Thomas J. Bridges}
\affiliation{School of Mathematics and Physics, University of Surrey, Guildford GU2 7XH, UK}

\begin{abstract}
For Stokes waves in finite depth within the neighbourhood of the Benjamin-Feir stability transition,  there are two families of periodic waves, 
one modulationally unstable and the other stable. In
this paper we show that these two families can be joined by a heteroclinic
connection, which manifests in the fluid as a travelling front. By shifting
the analysis to the setting
of Whitham modulation theory, this front is in
wavenumber and frequency space. An implication of this jump
is that a permanent frequency downshift of the Stokes wave can occur in the absence of viscous effects. This argument, which is built on a sequence
of asymptotic expansions of the phase dynamics,
 is confirmed via energetic arguments, with additional corroboration obtained by numerical simulations of a reduced model based on the Benney-Roskes equation.
\end{abstract}


\maketitle
\section{Introduction}
One of the most celebrated instabilities in fluid dynamics is the
Benjamin-Feir instability, where a Stokes wavetrain, travelling uniformly in
finite depth, undergoes a transition from stability to instability
as the depth of the fluid increases. The original papers \citep{bf67,b67}
have attracted significant attention in the years since its first publication. 
Nevertheless it continues to fascinate, and there is still much
to learn about its implications. 
One such phenomena, known now as frequency downshifting, emerged in experiments following up on the Benjamin-Feir result. 
These experimental investigations of monochromatic wavetrains  (including \cite{l77,sb82,m82,h96}) demonstrated that energy is exchanged from the primary wave mode to other sideband frequencies.
As this process begins to arrest, these experiments observed that the dominant peak of the wave power occurred not at the original carrier wave frequency, but one of a lower frequency, namely that the frequency peak had moved down the spectrum to lower frequencies (and thus the nomenclature).
In this paper we focus on modulation and dynamics of water waves near the Benjamin-Feir transition, and find that the frequency downshifting phenomena emerges naturally via phase dynamics.

The conventional explanation as to how this phenomena emerges is via dissipative effects,
such as wind forcing or inherent viscous effects, and these explanations have been supported by numerical simulations~\citep{lm85,hm91,dk99,cg16,ch19}.
It was thought that permanent frequency downshifting was not possible in purely conservative systems~\citep{lm85,hm91,dk99}.
However, as with all nonlinear paradigms, more than one mechanism can lead to the same phenomena.  There is a growing consensus that frequency
downshifting can indeed be observed without  energy dissipation or forcing~\citep{c07,c12,dtk03,j03,oos02,sks19}.
Whilst some of these alternative mechanisms have been observed numerically, 
an open question remains as to the theoretical explanation for downshifting to occur without dissipation. 

In this paper we propose a new mechanism for frequency downshifting for
inviscid and irrotational water waves without dissipation. 
The theory is based on asymptotically
valid modulation equations, building on classical Whitham modulation theory and its
generalisations. Whitham modulation theory has a distinct advantage over nonlinear
Schr\"odinger equation models (e.g. \cite{ho72,j77,km83}) in that it is precisely
the wavenumber and frequency that are modulated, thereby generating equations that inherently contain jumps in frequency. In the conservative setting, it is
singularities that provide the mechanism for downshifting. The
primary singularity is coalescence of two characteristics
in the Whitham modulation equations at the Benjamin-Feir transition \citep{w67}.

One has to go beyond \cite{w67} as higher-order modulation equations
are required in order to capture the nonlinear
implications of the double characteristic, which is what we achieve here within this paper. 
Our strategy is to re-scale and re-modulate to obtain new asymptotically valid modulation equations near the Benjamin-Feir threshold.  
In \cite{br17,br21} a general theory for the re-modulation of Whitham
theory in the neighbourhood of coalescing characteristics is constructed.
There it is found that the
conservation of wave action in Whitham theory is instead replaced by a two-way Boussinesq equation for
the modulation wavenumber, with the modulation frequency coming in via the equation for conservation of waves. However, this theory needs modification as a
secondary singularity arises at the Benjamin-Feir transition,
changing the nonlinearity in the two-way Boussinesq equation from quadratic to cubic. The resulting modulation equation, first derived in \cite{ratliff-thesis}, is
\begin{equation}\label{ratliff-eqn}
\alpha_1U_{TT} +\alpha_2(U^3)_{XX}+\alpha_3(2UU_T+U_X\partial_X^{-1}U_T)_X + \alpha_4 U_{XXXX} = 0\,,
\end{equation}
where $U$ characterises the local wavenumber, $T,X$ are slow
time and space scales, and $\alpha_1,\ldots,\alpha_4$
are real-valued parameters.  It is the key equation in this paper, as
it is asymptotically valid and contains travelling fronts which connect
 two wavenumber states thereby capturing the frequency downshifting via
conservation of waves.  The properties and analysis of (\ref{ratliff-eqn})
are given in \S\ref{sec-bf-transition-modulation}.

There are several steps in the analysis leading from the generic Whitham theory
to (\ref{ratliff-eqn}). The first step is to introduce 
a general form for the secondary 
modulation of the Stokes wave and meanflow.  The form of the
phase, wavenumber, and frequency modulation is cast in vector form as
\begin{equation}\label{alpha-beta-modulation}
\begin{gathered}
    \begin{pmatrix}
    \theta\\
    \phi_0
 \end{pmatrix}     =    \begin{pmatrix}
    k_0x-\omega_0t\\
    u_0x-\gamma_0t
 \end{pmatrix}  +\eps^\alpha {\bm \Theta}(X,T)\,, \\[3mm]
  \begin{pmatrix}
 k\\
 u
\end{pmatrix}   =  \begin{pmatrix}
k_0\\
u_0
\end{pmatrix}+\eps^{\alpha+1}{\bm K}(X,T)\,,\qquad
\begin{pmatrix}
\omega\\
\gamma
\end{pmatrix} =  \begin{pmatrix}
\omega_0\\
\gamma_0
\end{pmatrix}+\eps^{\alpha+1} c {\bf K}+\eps^{\alpha +\beta}{\bm \Omega}(X,T)\,,\\[3mm]
    {\rm with} \quad X = \eps (x-ct)\,, \quad T = \eps^\beta t\,, \quad {\rm and} \quad \eps \ll 1\,,
    \end{gathered}
\end{equation}
where $k_0,\omega_0$ are the wavenumber and frequency of
the Stokes wave, $u_0,\gamma_0$ are the bulk velocity and Bernoulli
constant of the meanflow.
The speed $c$ is a characteristic speed obtained from the generic
 Whitham modulation equations using the standard approach\citep{wlnlw}. The quantity $\theta$ is the phase of the wave, and $\phi_0$ is referred to as the pseudo-phase of the meanflow, due to its resemblance to a wave phase and playing a similar role in the Whitham Modulation equation.
The exponents $\alpha$ and $\beta$ are determined by assuring that
the equations are asymptotically valid, and that the modulation wavenumber and
frequences are in balance,
\[
 {\bm \Theta}_X = {\bm K} \quad {\rm and} \quad {\bm \Theta}_T = -{\bm \Omega}\,.
\]

We will look at two cases of the re-modulation.  The first is with
scales $\alpha=1$ and $\beta=3$.  These values are relevant in the hyperbolic
region, where all four characteristics in \cite{w67} are real, and
\[
c = c_g \pm \sqrt{\omega_0''(k_0) \omega_2^{eff}(k_0)} a+\ldots\,,
\]
with $\omega_0''\omega_2^{eff}>0$, where $\omega_2^{eff}$ is the
frequency correction to the Stokes wave, including meanflow.
  In this region we find that re-modulation leads to Korteweg de-Vries (KdV) dynamics, with the modulation equation taking the universal form~\citep{r21}
\begin{equation}\label{kdv-hyperbolic-region}
\Delta'(c) \bigg[U_T+\kappa UU_X+\frac{1}{6}\sigma'''(0)U_{XXX}\bigg] = 0\,,
\end{equation}
with $U$ characterising the evolution of the vector valued wavenumber via ${\bm K} = \zeta U(X,T)$ and $\zeta$ is the right eigenvector of the Whitham Modulation equations. Thus, the remodulation slaves the slow evolution of the wave and mean flow to one another within the water wave problem. The new modulation dynamics are characterised by properties of the wavetrain via the characteristic polynomial of the Whitham Modulation equations $\Delta(c)$ and the Bloch spectrum of the wave $\sigma(\nu)$ ~\citep{dsss09}.
The coefficient of the quadratic nonlinearity is,
\[
\kappa =  \begin{pmatrix}
\D_\bk c(\bk,\bw)\\
\D_\bw c(\bk,\bw)
\end{pmatrix} \cdot \begin{pmatrix}
\be\\
c \be
\end{pmatrix}\,,
\]
where $c(\bk,\bw)$ is a modulation (characteristic) speed and $\D$ denotes a directional (Gateaux) derivative,
can be interpreted as the linearised version of Lax's genuine nonlinearity criterion for the Whitham Modulation equations
\citep{r21}. The analysis leading to this equation, as
well as the definitions of $\Delta(c)$ and Bloch spectrum $\sigma(\nu)$, 
are given in \S\ref{sec:PD}.

It is important to note that the KdV equation (\ref{kdv-hyperbolic-region})
is not the famous KdV equation in shallow water hydrodynamics~\citep{kdv95}!  
It is a KdV equation describing perturbations of the Stokes wave and the meanflow, and not just long wave perturbations to the free surface and velocity (i.e. just the meanflow in the ansence of a background wave). 
This KdV equation is of interest here because in the limit to
the Benjamin-Feir transition, the coefficient $\kappa$ of the quadratic
nonlinearity goes to zero, signalling the change from quadratic to
cubic nonlinearity. This KdV equation may also have independent
interest in giving an alternative explanation for the appearance
of dark solitary waves in shallow water hydrodynamics (cf.\ \cite{bd06}).


In summary, the argument for frequency downshifting takes three steps.  Firstly,
generic Whitham modulation theory gives the characteristics with two of
these changing type from hyperbolic to elliptic at the BF transition.  
Secondly, remodulation in the hyperbolic region generates KdV dynamics on top of
the Stokes wave and meanflow.  Taking the limit to the BF transition then leads to 
a third modulation equation (\ref{ratliff-eqn}) for the wavenumber, and its
jump solutions generate frequency downshifting (or, in principle, upshifting). 
Whilst this paper will focus on the water wave problem as the key application of the above abstract theory, the
theory is more general as it applies to a basic
periodic wavetrain of any amplitude, as long as it has at least two phases.
The secondary modulation then is applicable.  This form of frequency downshifting
is universal in that the theory is formulated independent of any particular
equation, as long as it is conservative and generated by a Lagrangian.  However,
in this paper the focus is on the Benjamin-Feir transition.

An outline of the paper is as follows.  In \S\ref{sec-stokes} the theory for re-modulation is set up and those aspects of \cite{w67}
that feed into the higher-order modulation equations are highlighted.
Then in Section \ref{sec:PD},
the KdV equation on a Stokes wave (\ref{kdv-hyperbolic-region}) is
derived and analysed.  It is valid everywhere in the hyperbolic region of
generic Whitham theory, and we will be interested in its behaviour near the
hyperbolic-elliptic transition which signals the onset of Benjamin-Feir instability.
In \S\ref{sec:PD}, we also
introduce the concept of Bloch spectrum which 
arises in the derivation of the dispersive term in both (\ref{kdv-hyperbolic-region}) and in (\ref{ratliff-eqn}). In \S\ref{sec-bf-transition-modulation} the
key properties of (\ref{ratliff-eqn}) are highlighted, and the analysis
leading to jumps in wavenumber and frequency are given.  Further support
for the new theory of frequency downshifted in presented in \S\ref{subsec-energetics} using energy arguments, and in \S\ref{sec:BR-sim} by direct simulation of the Benney-Roskes equation. We find that the downshifted wavetrain is the state with the lower energy, providing an energetic argument for why downshifting is observed and persistent even in conservative systems. In the concluding remarks
section we summarise the main result, and indicate some generalisations.

\section{Stokes Waves, modulation and characteristics}
\label{sec-stokes}

In this section, we set up the basic state and its properties.
The basic state is a Stokes waves on finite depth coupled to meanflow.
The starting point for the analysis is
the inviscid, irrotational model for gravity waves in finite depth $h_0$
 and constant density.  The governing equations for the velocity potential $\phi(x,y,t)$,
and the free surface deflection $\eta(x,t)$ are
\begin{subequations}\label{WW-prob}
\begin{align}
\phi_{xx}+\phi_{yy} &= 0\,,\quad \mbox{for}\ y \in (-h_0,\eta)\,,\\[3mm]
\phi_y(x,-h_0,t) &= 0\,,\\[3mm]
\eta_t+\phi_x\eta_x &= \phi_y\,,\quad \mbox{at}\ y = \eta\,,\\[3mm]
\phi_t+\frac{1}{2}|\nabla\phi|^2+g \eta &= 0\,, \quad \mbox{at}\ y = \eta\,,
\end{align}
\end{subequations}
where $g$ is the acceleration due to gravity. These equations
are conservative and can be obtained from the first variation of a Lagrangian
\begin{equation}\label{Luke-Lag}
\delta L =0\,,\quad \mbox{with}\quad L= \int\int\bigg(\int_{-h_0}^\eta \phi_t+\frac{1}{2}|\nabla \phi|^2+g y\ dy\bigg)\ dx\,dt\,.
\end{equation}
The evaluation of $\delta L=0$ is given in \S13.2 of \cite{wlnlw}.
Now consider a Stokes expansion for the velocity potential and
 free surface, including an independent bulk mode represented by $b$,
\[
\begin{gathered}
\eta(\theta) = b+a \cos(\theta)+\sum_{n=2}^\infty a_n\cos(n \theta)\,, \\
 \phi(y,\theta) = \Phi+\sum_{n=1}^\infty\frac{A_n}{n}\cosh\big(nk(y+h_0)\big)\sin(n\theta)\,,
\end{gathered}
\]
with phases $\theta$ and $\Phi$ given by
\[
  \theta = kx-\omega t\qand \Phi = ux-\gamma t\,.
\]
The constants $k,\,\omega$ (representing the wavenumber and frequency) and $u,\,\gamma$ (representing the horizontal fluid velocity and Bernoulli head) parameterise the wave and the mean flow respectively. 

The quantity $\theta$ is the usual phase of the wavetrain, whereas $\Phi$ is called a pseudo-phase, although
mathematically it is equivalent to the wave phase (cf.\ \S3.1 in \cite{br21}). 
The pseudo-phase arises from an affine symmetry in $\phi$ present in the Lagrangian. 
The above solution, therefore, can be treated as a relative equilibrium with two phases -- one associated to the translation invariance in phase
of the Lagrangian, and the other due to the affine symmetry of the 
velocity potential.

Substitution  of the above wave-meanflow solution into the Lagrangian, averaging over one period of the wave, and solving the resulting system of
equations for the Fourier coefficients $a_n$ and $A_n$, one is able to obtain the following averaged Lagrangian, to leading
order in $E$ and $b$, 
\begin{equation}\label{Lag-WW}
\mathscr{L} = \bigg(\frac{u^2}{2}-\gamma\bigg)(h_0+b)+\frac{1}{2}g b^2+D(k,u,\omega)  E+\mu E \ b +\frac{1}{2}\tau E^2+\mathcal{O}(b^2,E^3, E^2 b)\,.
\end{equation}
The two small parameters are the energy density $E = \frac{1}{2}g a^2$ and mean
 deflection $b$ from the quiescent position $\eta = 0$.
 This reduced Lagrangian is derived in \cite{w67} and has been confirmed in \cite{br22}.

The function $D$ is the right-moving linear dispersion relation with a mean flow component,
\[ 
D(k,u,\omega) = \frac{1}{2}\bigg(1-\frac{(\omega-u k)^2}{\omega_0^2} \bigg)\,, \quad {\rm where} \quad \omega_0^2 = gk \tanh(kh_0)\,.
\]
It has a root at $\omega = uk+\omega_0(k)$.
 The constants $\mu$ and $\tau$ in (\ref{Lag-WW}) are
\begin{equation}\label{mu-tau-def}
\mu = \frac{B_0}{c_0h_0} \qand \tau = \frac{k^2}{g}\left(\frac{9T_0^4-10T_0^2+9}{8T_0^4}\right)\,,
\end{equation}
with
\[
B_0 = c_g-\frac{c_0}{2}\,, \quad c_g = u+\omega_0'(k)\,,\quad c_0 = \frac{\omega_0}{k}\,,
\quad\mbox{and}\quad T_0 = \tanh(kh_0)\,.
\]
Primes represent derivatives with respect to the wavenumber $k$. Variations of the Lagrangian (\ref{Lag-WW}) with respect to $E$ and $b$, when set to zero, yield the weakly nonlinear dispersion relations
\begin{equation}\label{non-DRs}
    D+\mu b+\tau E = 0\,, \qquad \gamma = \frac{u^2}{2} +g b+ \mu E
\end{equation}
In the absence of mean variations (i.e. $b=0$), the first can be solved to find the conventional low amplitude Stokes expansion of the frequency:
\[
\omega = uk+\omega_0(k)+\omega_2^0 E+\mathcal{O}(E^2)
\]
where the Stokes frequency correction in the absence of bulk/mean flow variations, $\omega_2^0$, is given by
\[
\omega_2^0 = \omega_0 \tau   = \frac{k^2 \omega_0}{g}\left(\frac{9T_0^4-10T_0^2+9}{8T_0^4}\right) = \frac{k^3}{\omega_0}\left(\frac{9T_0^4-10T_0^2+9}{8T_0^3}\right) = \frac{k^2}{c_0}\Lambda\,,
\]
where the expression $\Lambda$ is precisely $D_0$ in \cite{w67}, but the notation has been altered to prevent confusion. The dependence of $\omega_2^0$ on $k$ is important
and will be retained here as derivatives of $\omega_2^0$ with respect to these
parameters appear in the analysis, at leading order. 
Therefore, it is convenient to establish $\tau = \omega_0^{-1} \omega_2^0$, and thus we may write the coupled system (\ref{non-DRs}) as
\begin{equation}\label{RE-defn}
\begin{pmatrix}
\mu& \tau\\
g & \mu
\end{pmatrix}
\begin{pmatrix}
b\\
E
\end{pmatrix} = 
\begin{pmatrix}
-D\\
 \gamma-\frac{u^2}{2}
\end{pmatrix}\,.
\end{equation}
Henceforth, all expressions and coefficients within the paper, unless explicitly stated otherwise, will be evaluated at $\omega = uk+\omega_0(k)$.
It is assumed henceforth that these equations are non-degenerate
\[
\Delta_{W} = \mu^2-g \tau =\frac{B_0^2}{c_0^2h_0^2}-\frac{g \omega_2^0}{\omega_0} \neq 0\,.
\]
Indeed, this expression is negative-definite for gravity waves (but may change sign in other water wave problems, such as
 when surface tension or variable density is
present). This equation
clearly demonstrates that the energy density $E$ and bulk variation $b$ are truly independent, and not constrained as previously suggested in \S16.9 of
\cite{wlnlw}.  The independence of $b$ and $E$ play an important role in the phase modulation of the
Stokes waves, when $E$ and $b$ are slowly varying functions. The precise leading order
effect of the mean velocity field on the wave component is given in Appendix \ref{app-ab}.
Further, this system prescribes $E$ and $b$ as functions of the wave and meanflow parameters $k,\,\omega,\,u$ and $\gamma$, which will be required for the phase dynamical reduction. 

\subsection{Modulating wave and mean flow}

In classical Whitham modulation theory \citep{w67},
applied to the wave mean flow problem,
the key parameters,
\begin{equation}\label{th-k-w-def}
\bth = \begin{pmatrix}
 \theta\\
 \Phi
\end{pmatrix}\,, \quad
\bk = \begin{pmatrix}
k\\u
\end{pmatrix}\,, \quad \bw = 
\begin{pmatrix}
\omega\\
\gamma
\end{pmatrix}\,,
\end{equation}
are allowed to be slowly-varying functions
\begin{equation}\label{W-geom-optics}
\begin{gathered}
    \bth \to \bth +\eps^{-1}{\bm \Theta}(X,T)\,, \quad \bk \to \bk+{\bm K}(X,T)\,, \quad  \bw \to \bw+ {\bm \Omega}(X,T)\,,\\[3mm]
    {\rm where} \quad X = \eps (x-ct)\,, \quad T = \eps t\,, \quad {\rm and} \quad \eps \ll 1\,.
    \end{gathered}
\end{equation}
The coupled Whitham modulation equations are then
\begin{equation}\label{multiphase-wmts}
{\bf K}_T = {\bm\Omega}_X \qand \frac{\partial\ }{\partial T}
{\bf A}(\bw + {\bm\Omega},\bk+{\bf K}) +
\frac{\partial\ }{\partial X}
{\bf B}(\bw + {\bm\Omega},\bk+{\bf K}) = 0 \,,
\end{equation}
(cf.\ equation (1.14) of \cite{br21}), where the first equation is the so-called "conservation of waves", and the second is the conservation of wave action for each phase. As such, ${\bf A}$ and ${\bf B}$ are denoted as
the vector valued wave action and wave action flux respectively,
\begin{equation}\label{AB-def}
{\bf A} = -\D_\bw\mathscr{L} \qand {\bf B} = \D_\bk\mathscr{L}\,.
\end{equation}
Differentiating $\mathscr{L}$ in (\ref{Lag-WW}) and substituting into
(\ref{AB-def}) gives the components of the wave action conservation law,
\begin{equation}\label{CLaws-WW}
\begin{gathered}
{\bf A} = -\D_\bw\mathscr{L} = 
\begin{pmatrix}
-D_\omega E\\
h_0+b
\end{pmatrix}\,, \quad {\bf B} =\D_\bk\mathscr{L} = 
\begin{pmatrix}
D_k E+\mu ' Eb+\frac{1}{2}\tau' E^2\\
u(h_0+b)+D_u E
\end{pmatrix}\,, \\[4mm]
 {\rm where} \quad (\D_{\bm x}F){\bm y} = \lim_{s\to0}\bigg(\frac{F({\bm x}+s{\bm y})-F({\bm x})}{s} \bigg)\,.
 \end{gathered}
\end{equation}
To compute characteristics, we will need the linearisation
of (\ref{multiphase-wmts}).  Differentiating (\ref{multiphase-wmts}) with
respect to ${\bm\Omega}$ and ${\bf K}$, linearising, and introducing the
characteristic form
\[
{\bm\Omega}(X,T) = \widehat{\bm\Omega}\re^{\ri(X+cT)}\qand
{\bf K}(X,T) = \widehat{\bf K}\re^{\ri(X+cT)}\,,
\]
results in an eigenvalue problem for the characteristics $c$,
\begin{equation}\label{wme-char-eqn}
\left[ \begin{pmatrix} -\D_\bw{\bf A} & {\bf 0} \\ {\bf 0} & \D_\bk{\bf B} \end{pmatrix}
+c \begin{pmatrix} {\bf 0} & \D_\bw{\bf A} \\\ D_\bw{\bf A} & \D_\bk{\bf A}+\D_\bw{\bf B}
\end{pmatrix}\right] \begin{pmatrix} \widehat{\bm\Omega}\\ \widehat{\bf K}\end{pmatrix} =
\begin{pmatrix} {\bf 0}\\ {\bf 0}\end{pmatrix}\,.
\end{equation}
This equation is an example of the general form of the equation for
characteristics in multiphase Whitham modulation theory (cf.\ equation (1.18) in \cite{br21}). It is assumed in this construction that $D_\bw{\bf A}$ is invertible, and
the first equation in (\ref{wme-char-eqn}) has been multiplied by this matrix.

Combining the two equations in (\ref{wme-char-eqn}) by eliminating
$\widehat{\bm\Omega}$, and defining $\be=\widehat{\bf K}$,
reduces this equation to the matrix pencil,
\begin{equation}\label{E-def}
{\bf E}(c)\be = 0\,,\quad \be = \begin{pmatrix}\zeta_1\\ \zeta_2\end{pmatrix}\,.
\end{equation}
The roots of ${\rm det}({\bf E}(c))=0$ are the characteristics of the modulation
equations for the Stokes wave meanflow interaction.
The general form of the $2\times 2$ matrix $E(c)$ is
\[
{\bf E}(c) = \D_\bk{\bf B}+c(\D_\bw{\bf B}-\D_\bk{\bf A})-c^2\D_\bw{\bf A}
:= \left(\begin{array}{cc} E_{11} & E_{12}\\  E_{12}&E_{22} \end{array}\right)\,,
\]
and explicit expressions for the entries $E_{11},E_{12}$ and $E_{22}$ for
the water-wave problem are given
in Appendix \ref{app-E-Expr}.

From this pencil, we find the characteristic polynomial
for the Stokes wave meanflow modulation is
\begin{equation}\label{char-poly-ww}
\begin{split}
\Delta(c) = &{\rm det}\big( {\bf E}(c)\big) \\[2mm]
=&(c-c_g)^2\big(gh_0-(u-c)^2\big)-2\omega_0(c-c_g) \big(gh_0-(c-u)^2\big)\mu'\, b \\[3mm]
& \hspace{1.0cm}-E \left\lbrace \omega_0''\Omega+2(c-c_g)\bigg[\omega_0(gh_0-(c-u)^2)\tau'-\frac{B_0+(c-u)}{\omega_0}\mu'\right.\\
&\hspace{5.0cm}\left.-\frac{\omega_0'}{\omega_0}\Omega+\frac{\big(g h_0+ B_0(c-u)\big)}{c_0h_0}\bigg(\frac{\omega_0'}{c_0}-1\bigg)\bigg]\right\rbrace\\[3mm]
&\hspace{1.0cm}+ \mathcal{O}(E^2,Eb,b^2,(c-c_g)^2E,(c-c_g)^2b)\,,
\end{split}
\end{equation}
where
\begin{equation}\label{char-poly-ww-1}
\Omega(c;k,u) =  \omega_2^0\big(gh_0-(c-u)^2\big)-\frac{k}{c_0 h_0}(B_0^2+2B_0(c-u)+gh_0)\,.
\end{equation}
This polynomial has four roots, admitting 4 characteristics, and we will find explicit
expressions for them in the small amplitude limit $E\ll 1$.  Two of these characteristics are associated with the group velocity of the wave, whereas the final two are related to the linear long wavespeeds. It is the former that we will be interested in, as these are the ones which correspond to the Benjamin-Feir instability. The two characteristics of (\ref{char-poly-ww}) that are
associated with the group velocity, for small amplitude, are found to be
\begin{equation}\label{char-ww}
\begin{split}
c = u+\omega_0' \pm \sqrt{\omega_0'' \omega_2^{eff}E}+C_2E+\omega_0\mu' b+\mathcal{O}(E^\frac{3}{2},Eb,b^2)\,,
\end{split}
\end{equation}
where
\[
\omega_2^{eff} =\frac{\Omega(c_g;k,u)}{gh_0-\omega_0'^2} = \frac{k}{c_0 h_0}\bigg[k h_0 \Lambda-\bigg(\frac{B_0^2+2B_0 \omega_0' +g h_0}{g h_0 -\omega_0'^2} \bigg)\bigg]\,,
\]
and we have defined for brevity
\[
\begin{split}
C_2 &= \big(\omega_2^0\big)' +\frac{B_0\omega_0'+gh_0}{c_0h_0(gh_0-\omega_0')}\bigg(\frac{k(B_0+\omega_0')}{gh_0-\omega_0'^2} \omega_0''-1\bigg)-\frac{(B_0+\omega_0')}{gh_0-\omega_0'^2}\frac{(kB_0)'}{c_0h_0}\,.\\
\end{split}
\]
This expression for $\omega_2^{eff}$ is what is called $\Omega_2(k)$ in \S16.11 in \cite{wlnlw}.
However,
the way it has emerged in this analysis is surprising, as the assumptions are
different.
In Whitham's analysis, one must either identify several terms in the analysis and modulation equations to neglect on consistency or size arguments (as in~\cite{w67}) or appeal to the flux induced by the waves via an argument proposed by Longuet-Higgins in order to arrive at the correct frequency expression (as done in~\cite{wlnlw}), which essentially constrains the mean induced by the waves $b$ to be related
to the energy density $E$. The relative equilibrium approach here, which maintains the
independent of $b$ and $E$, does not require this additional information and suggests that
the result in Whitham is in fact a consequence of the symmetries of the Lagrangian and thus inherent to the problem itself. 

We will also need the eigenvector of ${\bf E}(c)$ when $c$ is
a characteristic,
in the asymptotic limit $E\to 0$.
In the neighbourhood of the double characteristic, the unfolding of the critical point
will be of $\mathcal{O}\big(E^{\frac{1}{2}}\big)$.  In the equation
${\bf E}(c)\be=0$ the characteristic (\ref{char-ww}) is substituted
in for $c$ and $\be$ is expanded in powers of $E^{\frac{1}{2}}$.
The details of this expansion can be found in (\ref{eigenvect})
 of Appendix \ref{app-E-Expr}. Evaluating this eigenvector is at
 the BF transition $kh_0 1.363$, it becomes
\[
\be=\sqrt{E}\left[C_2{\bm \chi}-\begin{pmatrix}
\frac{\omega_0'}{c_0}-1+\frac{1}{\Delta_W}\big(\omega_2^0\omega_0'+\frac{kB_0}{c_0h_0}\big)\mu'-\frac{k(B_0\omega_0' + gh_0)}{h_0\Delta_W}\tau'\\
 \omega_0''
\end{pmatrix} \right]+ \mathcal{O}(E^{\frac{3}{2}})\,,
\]
with
\[
{\bm \chi} = -\frac{1}{c_0h_0\Delta_W} \begin{pmatrix}
gh_0+B_0\omega_0'\\ 0 
\end{pmatrix}\,.
\]
This eigenvector
evaluated at the transition value simplifies to
\[
\be =  \begin{pmatrix}
0.9252 \\
2.1703h_0^\frac{3}{2}
\end{pmatrix}\, a + \mathcal{O}(a^3)\,,
\]
since $E=\frac{1}{2} g a^2$.

\subsection{Bloch Spectrum}
\label{sec:Bloch}
As can be see from (\ref{ratliff-eqn}) and (\ref{kdv-hyperbolic-region}), a key component of the phase dynamical construction associated to the coefficient of dispersion in the problem is the
 Bloch spectrum $\sigma(\nu)$, with $\nu$ the spatial Floquet exponent/Bloch wavenumber, for the Stokes wave solution. 
The Bloch spectrum consists of the eigenvalues of the linearisation of the full water-wave problem about
the Stokes wave and has a central place in understanding the stability of Stokes waves in finite depth \citep{do11,cd23,bmv23,bmv24}, and so it is unsurprising that it features as a component of the phase dynamics reduction.
In this paper only third order dispersive effects are required in the asymptotic
analysis because only the third order Taylor coefficient of the Bloch spectrum about the zero Bloch wavenumber is required. In this section we will identify an appropriate choice for the Bloch spectrum that is both analytically tractable and representative of the problem.

Due to the low amplitude nature of the Stokes waves within this paper, it would be fair to expect the spectrum that arises from such problems to be akin to the spectrum of NLS-type models:
\begin{equation}\label{sigma_HONLS}
\sigma_{HNLS}(\nu) =c_g \nu+ \frac{\omega_0'''}{6}\nu^3 \pm \nu \sqrt{\omega_0''\omega_2^{eff}E+\frac{1}{4}\omega_0''^2\nu^2}\,,
\end{equation}
where the subscript $HNLS$ indicates that this is the exact spectrum for the
NLS equation with a higher order dispersion terms \citep{r21}.
To leading order in $E$ and without higher order disperive effects, this expansion has been shown the same in the full water wave problem in the vicinity of the Benjamin-Feir instability~\citep{bm95}. However the mean flow effects in NLS models are treated adiabatically and as such may not be captured fully in this Bloch spectrum. 
To remedy this we compare this NLS Bloch spectrum to a model where the mean flow is nonadiabatic, which within this paper will be the one obtained from the Benney-Roskes equation, given by~\citep{br69}
\begin{gather}
iA_T+\epsilon \bigg(A_{XX}-\omega_2|A|^2-k WA-\frac{gk^2-\omega_0^4}{2g\omega_0}BA\bigg) = 0\label{BR-1}\\
B_T-c_gB_X+h_0W_X+\frac{gk^2-\omega_0^4}{\omega_0^2}(|A|)_X^2 = 0\label{BR-2}\\
W_T-c_gW_X+gB_X+\frac{2gk}{\omega_0}(|A|^2)_X = 0\label{BR-3}\,.
\end{gather}
In this equation, $A$ is the complex amplitude of the wave, $B$
is the mean level variation, and $W$ is the mean velocity.
The small parameter $\epsilon$ (which differs from $\eps$ in the modulation
theory) gives the order of magnitude of the wave amplitude.  It is equivalent to the smallness of the parameter $a$ and thus allows one to relate the Bloch spectrum of the Benney-Roskes system to the modulation theory of this paper.
Whilst the full closed form expression for the spectrum is complicated, we only require its long wave expansion, which in terms of the Stokes wave parameters reads
\[
\sigma_{BR}(\nu) = c \nu\pm\bigg(\frac{\omega_0''^2}{8 \sqrt{\omega_0'' \omega_2^{eff}E}}+\Xi\bigg) \nu^3+\mathcal{O}(\nu^5)\,,
\]
where the characteristic $c$ is defined as in (\ref{char-ww}) and
\[
\Xi = \frac{g\omega_0''^3k((3c_g^2 + g h_0)(B_0^2 + 2 B_0c_g + g h_0) + 4 B_0 c_g (gh_0-c_g^2))}{4c_0h_0(gh_0-c_g^2)^2\sqrt{\omega_0''\omega_2^{eff}}}\sqrt{E}+\mathcal{O}(E)\,,
\]
characterises the additional dispersive effects due to the mean flow.
We note that this spectrum only differs from the long wave expansion of the classical Nonlinear Schr\"odinger equation at $\mathcal{O}(\sqrt{E})$, but does not alter the Benjamin-Feir instability boundary.  As such, we postulate that we may augment the above spectrum with the third order dispersive term in (\ref{sigma_HONLS})\footnote{Formally, this can be done by following asymptotic procedures such as in \cite{s05} or \cite{km83}, noting that the mean flow effects on the stability have already been accounted for.} to give the Bloch spectrum we shall utilise within this paper:
\begin{equation}\label{Spectrum}
\sigma(\nu) = c \nu+\bigg(\pm\frac{\omega_0''^2}{8 \sqrt{\omega_0'' \omega_2^{eff}E}} \pm \Xi+\frac{\omega_0'''}{6}\bigg) \nu^3+\mathcal{O}(\nu^5)\,.
\end{equation}

\section{Phase modulation in the hyperbolic region}
\label{sec:PD}

The hyperbolic region is the region in which Stokes waves coupled to meanflow
exist and the Whitham modulation equations, linearised about the
Stokes wave (\ref{wme-char-eqn}), have four real characteristics. 
Everywhere in this region the modulation can be re-scaled as in (\ref{alpha-beta-modulation})
to derive the KdV equation (\ref{kdv-hyperbolic-region}).  This KdV equation is in
a characteristic moving frame with the speed determined by a characteristic from
the generic Whitham theory. The theory for this remodulation follows \cite{rb16} and
\cite{r19,r21}. 
The form for the remodulation is (\ref{alpha-beta-modulation}) with $\alpha=1$ and
$\beta=3$.  The velocity potential and free surface are expressed as
\[
Z(x,y,t) = \Zh\big({\bm\theta}+\eps{\bm\Theta}(X,T);{\bf k}+\eps^2{\bf K}(X,T),{\bm\omega} + \eps^2 c{\bf K} +\eps^3{\bm\Omega}\big) +\eps^4 W(X,T,\eps)\,.
\]
where $Z(x,y,t)=(\phi(x,y,t),\eta(x,t))$, $\Zh({\bm\theta};{\bf k},{\bm\omega})$ 
is the Stokes wave plus mean flow, and $c$ is one of the wave characteristics
in the hyperbolic region,
\begin{equation}\label{char-hyperbolic}
c = c_g \pm \sqrt{\omega_0''(k_0) \omega_2^{eff}(k_0)} a+\ldots\quad\mbox{with}
\quad \omega_0''(k_0) \omega_2^{eff}(k_0)>0\,.
\end{equation}
The modulation equations are then obtained by substitution of the above
 Stokes wave solution with these perturbed wave quantities into the water wave equations and solving the resulting system at each order of the small parameter $\eps$. The
strategy is given in \cite{rb16} and \cite{r19} and so we skip details. The resulting
KdV equation has the form given in (\ref{kdv-hyperbolic-region}), which
we repeat here as we will evaluate the key coefficients,
\begin{equation}\label{kdv-hyperbolic-II}
\Delta'(c) \bigg[U_T+\kappa UU_X+\frac{1}{6}\sigma'''(0)U_{XXX}\bigg] = 0\,.
\end{equation}
The function $U(X,T)$ in (\ref{kdv-hyperbolic-II})
is obtained by projection of ${\bf K}(X,T)$ in the
direction of the eigenvector $\be$ of ${\bf E}(c)\be=0$ with ${\bf E}(c)$ defined
in (\ref{E-def}) with its argument evaluated at (\ref{char-hyperbolic}),
\begin{equation}\label{K-be-def}
{\bf K}(X,T) = U(X,T)\be\,.
\end{equation}
It is important to note that this KdV equation is not the classical
KdV equation in shallow water, which can also be derived using
phase dynamics \citep{b14}, but the coefficients and implications are different. This is because in addition to perturbing the mean free surface level and horizontal velocity that the classical KdV would suggest, it additionally alters the wavenumber, frequency and amplitude of the surface Stokes wavetrain with nonzero amplitude. 
Indeed, the solitary wave solution of (\ref{kdv-hyperbolic-region}) is in fact a dark solitary wave (bi-asymptotic to a Stokes travelling wave).  Hence the hyperbolic region is not only filled with modulationally stable Stokes waves, it is also filled with dark solitary waves, each moving at its local characteristic speed (\ref{char-hyperbolic}).

The coefficient $\Delta(c)$ is defined in (\ref{char-poly-ww}), and its
derivative is found to be 
\begin{equation}\label{delta-prime-kdv}
\Delta'(c) = \pm 2\big(g h_0-\omega_0'^2\big)\sqrt{\omega_0'' \omega_2^{eff}E}+\mathcal{O}(E^{\frac{3}{2}})\,.
\end{equation}
The second important term is the dispersive term $\sigma'''(0)$ in the KdV equation, which is obtained from the Bloch spectrum.  The Bloch spectrum is the temporal eigenvalue
$\sigma(\nu)$ of the linearisation of the full equations considered as a function
of the spatial Floquet exponent (see also \S\ref{sec:Bloch} below).
Using the spectrum (\ref{Spectrum}), we can readily obtain
\[
\begin{split}
 \frac{1}{6} \sigma'''(0) =& \pm \frac{\omega_0''^2}{8 \sqrt{\omega_0'' \omega_2^{eff}E}}\pm \Xi+\frac{\omega_0'''}{6}+\mathcal{O}(E).\,.
\end{split}
\]
Note that although this expression appears to be singular at the Benjamin-Feir transition $\omega_0''\omega_2^{eff} = 0$ the singularity is of the same order as the zero of $\Delta'(c)$ at this point, meaning the dispersive term in the KdV is finite and nonzero at this transition.

The third coefficient of interest is the coefficient $\kappa$ of the nonlinearity
\[
\kappa = \begin{pmatrix}
\D_\bk c(\bk,\bw)\\
\D_\bw c(\bk,\bw)
\end{pmatrix} \cdot \begin{pmatrix}
\be\\
c \be
\end{pmatrix}\,.
\]
This latter expression is related to the concept of
``genuine nonlinearity'', of the Whitham modulation equations, in the sense of Lax (cf.\ \cite{l73} and \cite{r21}). Evaluation of this coefficient for the water-wave problem
is straightforward but lengthy. Using the expressions (\ref{char-ww})
 and (\ref{eigenvect}) we can show that
\begin{equation}\label{kappa-defn}
\begin{split}
\kappa =& \mp\frac{3}{2}\mathscr{M}_1\sqrt{\omega_0'' \omega_2^{eff}}-\sqrt{E}\left\lbrace \omega_0''\bigg[\frac{\omega_0'}{c_0}+\frac{\mathscr{M}_1}{2}\big((\omega_2^{eff})'+4C_2-2\tau'\omega_0\big)\bigg]\right.\\
&\hspace{1cm}\left. +\omega_2^{eff}\bigg[\frac{1}{2}\mathscr{M}_1\omega_0''' +\omega_0''\bigg(\frac{3(\omega_0'\mu)'}{2\Delta_W}+\frac{\omega_0'}{\omega_0}\mathscr{M}_1-\frac{g(k\omega_0'-3\omega_0)}{2\omega_0^2\Delta_W}\bigg)\bigg]\right\rbrace+\mathcal{O}(E)\,.
\end{split}
\end{equation}
The coefficient $\mathscr{M}_1$ is given in equation (\ref{M-defns})
of Appendix \ref{app-ab} where it is associated with the change of wave properties
due to mean velocity changes.
The coefficient of the nonlinearity $\kappa$ is finite at the Benjamin-Feir transition, and so
once multiplied by (\ref{delta-prime-kdv}) the quadratic term in the KdV will vanish. This
will be important at the BF transition, and is the reason that the nonlinearity within the phase dynamical description goes
from purely quadratic as in (\ref{kdv-hyperbolic-region}) to involving the cubic and mixed quadratic terms seen in(\ref{ratliff-eqn}).

In summary, in the BF stable region, the KdV equation
 which emerges, for the faster of the two
characteristic speeds, is
\begin{equation}\label{kdv-SW}
\begin{split}
&\sqrt{\omega_0'' \omega_2^{eff}E}\,(U_T+\kappa UU_X)-\bigg[\frac{\omega_0''^2}{8}+\bigg(\Xi+\frac{\omega_0'''}{6}\bigg)\sqrt{\omega_0'' \omega_2^{eff} E}\bigg]U_{XXX} = 0
\end{split}
\end{equation}
This KdV equation is asymptotically correct to order $E$.  
However, the KdV equation does not support heteroclinic connections, thereby precluding jumps in frequency and wavenumber. Hence, in the strictly hyperbolic (modulationally stable) regime 
there cannot be a permanent change in wavenumber. 

As noted above, it is apparent from the expressions for the coefficients that that the first two terms of this KdV are zero whenever $\omega_2^{eff} = 0$,  occurring exactly at the Benjamin-Feir transition, which signifies a change in scale. This inevitably leads to at least cubic nonlinearities emerging, although quadratic nonlinearities of mixed type (i.e., involving spatial and temporal derivatives) are {\it a priori} also anticipated.  This theory is developed in the next section, resulting in a modified version of the two-way Boussinesq equation.  This new modulation equation
will have the necessary nonlinearities and dispersion to support a heteroclinic connection that will lead to downshifting of the Stokes waves.

\section{Phase modulation near the Benjamin Feir Transition}
\label{sec-bf-transition-modulation}

In approaching the Benjamin-Feir transition, two singularities arise.  Firstly,
two characteristics coalesce as noted in \S\ref{sec-stokes} and secondly, the coefficient of
the quadratic nonlinearity vanishes as noted in \S\ref{sec:PD}. In light of this we utilise time scaling typically used to derive two-way Boussinesq equations, used to rebalance the time portion of the dynamics~\citep{rb16a}, in tandem with scalings used to obtain the modified KdV equation where cubic terms resolve vanishing quadratic nonlinearities \citep{gg83,r21}. Therefore in the re-modulation 
the coalescing characteristics change in light of the above observations, the first resulting in the small time exponent changing from $\beta=3$ to $\beta=2$ and the second a change of perturbation scales from $\alpha = 1$ to $\alpha = 0$.  With
the loss of quadratic nonlinearity, and emergence of cubic nonlinearity new terms appear in
the equation as shown in (\ref{ratliff-eqn}) rewritten here in a different form
\begin{equation}\label{ratliff-eqn-1}
\alpha_1U_{TT} + \big(\alpha_2U^3 + \alpha_4 U_{XX}\big)_{XX} 
+\alpha_3(2UU_T+U_X\partial_X^{-1}U_T)_X =0\,.
\end{equation}
The first three terms are the two-way Boussinesq equation with a cubic nonlinearity.
The latter term, multiplied by $\alpha_3$, is required to balance the cubic nonlinearity.
With $U$ of order $\eps$, $X$ of order $\eps$ and $T$ of order $\eps^2$, the three nonlinear terms
are in balance
\[
(U^3)_{XX} \sim \eps^5\,,\quad (UU_T)_X \sim \eps^5\,,\qand (U_X\partial_X^{-1}U_T)_X \sim \eps^5\,.
\]
A detailed derivation of this equation is given in \S4.5.3 of \cite{ratliff-thesis} for the case of the
laboratory frame, but can be extended to the case of the characteristic moving frame using recent works
by the authors (most notably, \citet{br17,r21}).
The dependent variable $U$ is again obtained as a projection of the wavenumber onto
the eigenvector, ${\bf K}=U\be$, as in (\ref{K-be-def}), although here the eigenvector is
the one associated with coalesced characteristics.

General expressions for the parameters in (\ref{ratliff-eqn-1})
are given in terms of the averaged Lagrangian in equation (4.27) of
\cite{ratliff-thesis}, however the more accessible way to compute this number is to use
the connection of the coefficients to the Bloch spectrum and expansions of the
flux vector as in \citet{r21}. The evaluation of the coefficients for the water wave problem at the
Benjamin-Feir transition are lengthy, and results are summarised here.
The dispersion coefficient is
\[
\alpha_4 = -\frac{\omega_0''^2}{8}(gh_0-\omega_0'^2)\,.
\]
The time derivative term, which goes from a first-order derivative
to a second-order derivaive at the BF transition has coefficient
\[
\alpha_1 = -\frac{1}{2}\Delta''(c) = \omega_0'^2-gh_0\,.
\]
The most complicated coefficient is that multiplying the cubic
nonlinearity. Evaluating the formula on the weakly nonlinear
Stokes wave gives
\[
\alpha_2 = \frac{1}{2}\left[\Delta'' \kappa^2+\kappa \left[ 
\begin{pmatrix}
\D_\bk \Delta'\\
\D_\bw \Delta'
\end{pmatrix} \cdot 
\begin{pmatrix}
\be\\
c \be
\end{pmatrix}\right]
\right] = -\frac{\kappa \omega_0''}{12\Delta_W}((\omega_2^{eff})'+4C_2)\,.
\]
At the Benjamin-Feir transition, $\kappa$ reduces to
\[
\begin{split}
\kappa =& -\sqrt{E} \omega_0''\bigg[\frac{\omega_0'}{c_0}+\frac{\mathscr{M}_1}{2}\big((\omega_2^{eff})'+4C_2-2\tau' \omega_0\big)\bigg]+\mathcal{O}(E)
\end{split}
\]
The last coefficient to compute is for the quadratic terms, which emerge due to the simultaneous vanishing of the time and nonlinear terms in the KdV term. This has the coefficient
\[
\alpha_3 = \Delta'' \kappa
\]
Overall, this gives the modified Two-way Boussinesq as
\begin{equation}\label{BF-eqtn}
\begin{array}{rl}
&U_{TT}+\beta_1(U^3)_{XX}+\beta_2(2UU_T+U_X\partial_X^{-1}U_T)+\beta_3 U_{XXXX} = 0\,,\\[5mm]
{\rm with} \qquad &\beta_1 = -\frac{\kappa\omega_0''}{6\Delta_W}\big[(\omega_2^{eff})_k+4C_2\big]\sqrt{E}\\[3mm]
&\beta_2 = -2\kappa\\[3mm]
&\beta_3 = \frac{\omega_0''^2}{4}\,.
\end{array}
\end{equation}
The remaining depth and amplitude effects, to leading order, can be removed with the rescaling
\[
T = \sqrt{\frac{h_0}{g}}\tau\,, \quad X = h_0 \chi\,, \quad U = a^{-1}h_0^{-\frac{3}{2}}\mathcal{U}\,,
\]
reducing the phase dynamical equation at the Benjamin-Feir transition point to simply
\begin{equation}\label{BF-nondim}
\mathcal{U}_{\tau \tau}-\big(0.1182\mathcal{U}^3-0.0333\mathcal{U}_{\chi \chi}\big)_{\chi \chi}+0.8208\big(2\mathcal{U}\mathcal{U}_\tau+\mathcal{U}_\chi\partial_\chi^{-1}\mathcal{U}_\tau\big)_\chi = 0\,.
\end{equation}
It will be this equation which we analyse in order to determine the evolution of the wave and mean flow at the Benjamin-Feir transition, and the downshift phenomenon.

\subsection{Heteroclinic connections representing frequency downshifting}
\label{sec:FD-sol}

We now solve (\ref{BF-nondim}), postulating travelling wave solutions of the form 
\[
\mathcal{U}(X,T) = R(\xi)\,,\quad\mbox{with}\quad \xi = X - V T\,,
\]
parameterised by $V$, where $V$ is the speed of the travelling front.
To capture permanent downshifting of the Stokes waves we will prescribe the boundary conditions
\[
\lim _{\xi \to \infty}R(\xi) = K_1\,, \quad \lim_{\xi \to -\infty} = K_2\,, \qquad K_2\neq K_1\,.
\]
These boundary conditions correspond to a pair of asymptotic wavenumbers for the perturbed Stokes waves of the form $k_{1,2} = k_0+\eps \zeta_1 K_{1,2}$, connecting initial state $k_1$ (assuming $V>0$ without loss of generality) to $k_2$. 
These assumptions transform (\ref{BF-nondim}) into an ODE, which may be integrated to form a system possessing a quartic potential, with the travelling front now represented by a heteroclinic
connection.

The boundary conditions impose that the quartic potential of the ODE
 must possess two repeated roots at $K_1$ and $K_2$, and so the system for the frequency downshifting solution takes the form
\begin{equation}
    \bigg(\frac{dR}{d \xi^2}\bigg)^2 +\frac{\beta_1}{2\beta_3}(R-K_1)^2(R-K_2)^2 = 0\,.
\end{equation}
A comparison between (\ref{BF-nondim}) and the above gives that the far-field states take one of the following pair of values
\[
K_{1,2} = \frac{V}{2\beta_1} \big(\beta_2 \pm \sqrt{3\beta_2^2-4\beta_1}\big) = \big(-3.4721 \pm 6.6803\big) V
\]
The square root exceeding the leading fraction ensures the conjugate states lie on opposite sides of the Benjamin-Feir threshold, and thus connect a Stokes wave which is modulationally unstable to one which is modulationally stable.
This is valuable insight, as this suggests that a transition which leads to $K_2>K_1$ 
would be inadmissible owing to the fact that the state it is attempting to connect to an unstable
wavetrain, rather than a uniform wavetrain. 
Thus, it follows that one should choose $K_2<0$ and $K_1>0$ on physical grounds to avoid such a scenario.  There is an alternative reasoning based on energetics that may also be employed, which is described below in \S\ref{subsec-energetics}.

Moreover,  a secondary insight is that the speed of this transition between biasymptotic states is linked to the size of the wavenumber transition, suggesting larger deviations from the carrier wave will be resolved much more rapidly, in line with what one would expect experimentally.
With these boundary conditions and reasoning,
 this double root corresponds to the jump profile
\begin{equation}\label{R-front}
R(\xi) = \frac{1}{2} \ \big[K_1+K_2+(K_1-K_2)\tanh(2.0067(K_2-K_1)\xi) \big]\,.
\end{equation}
The solution family presented here accords novel insight into the
frequency downshifting phenomena from a conservative but dispersive
point of view.  On the other hand, it is useful to discuss it's limitations.  Primarily, the solution here presents the connection between two biasymptotic states but is unlikely to accurately describe the evolution of the wave as it transitions between them. This is due to the fact that a great number of the higher harmonics and their sideband contribute to the energy transfer within the wavetrain. This is apparent in the original experiments of \cite{l77}, where there is a devolution from sideband and harmonic dynamics to a much broader spectral wave evolution.

\section{Energetics of Frequency Downshifting}
\label{subsec-energetics}

This analysis of the previous section highlights that two heteroclinic connections are supported by this system,  which initially suggests both upshifting and downshifting are permissible.  Here we provide some discussion as to how this can be interpreted energetically, leading us to conclude that
frequency downshifting arises instead of upshifting.  In this discussion, we will denote the connection where $k_2<k_0$ as the lower sideband solution and $k_1>k_0$ as the upper sideband.  These correspond to the biasymptotic states of $U$ characterised by $K_2<0<K_1$ respectively.

We begin our discussion with the energy density of the wave, $E$, under the action of the jump solution (\ref{R-front}). By comparing lower and upper sideband wavenumbers we are able to show from (\ref{LO-perts}) that the energy density of the sidebands is related to the energy density of the carrier wave, $E_0$, to leading order via
\[
E_{1,2} = E_0-\eps \sqrt{E}\,\omega_0''\mathscr{M}_1K_{1,2}+\mathcal{O}(\eps^2 \sqrt{E},\eps E)\,.
\]
Thus, it is clear that $E_1<E_0<E_2$ and so more energy is passed to the lower sideband than the upper sideband under the jump mechanism. This is in line with experimental (e.g.\ \cite{l77,m82}) and theoretical observations (e.g.\ \cite{b82}). The primary driver of this energy exchange is the mean flow effect, suggesting the mechanism for the sideband asymmetry is indeed a mean flow aspect of the problem rather than the wave. Additionally, the energy of the lower sideband as $T\to \infty$ under this mechanism exceeds that of the carrier wave, as also seen in the aforementioned studies.  These facts together suggest that there is an overall shift in energy downwards in the spectrum over long time, and thus the spectral peak moves from the carrier wavenumber to that of the lower sideband.  

The energy density alone does not, however, indicate whether the upshift or downshift is ultimately selected by the system but can be resolved by looking at the total wave energy.  Recall the definition of wave energy for the water wave problem~\citep{wlnlw}:
\[
\mathscr{E} = \frac{1}{2}(h_0+b)\bigg(u+\frac{E}{c_0(h_0+b)}\bigg)^2+\frac{1}{2}g (h_0+b)^2+E\,.
\]
Let us denote the wave energy of the carrier wave by $\mathscr{E}_0$, then to leading order the energy of the upper and lower sidebands are
\[
\mathscr{E}_{1,2} = \mathscr{E}_0 -\eps \sqrt{E}\omega_0''K_{1,2} \bigg[gh_0\mathscr{M}_2+\mathscr{M}_1\bigg]+\mathcal{O}(\eps E, \eps^2 \sqrt{E},\eps b)\,,
\]
with $\mathscr{M}_2$ defined in equation (\ref{M-defns})
of Appendix \ref{app-ab}.
It follows from evaluating the above at the Benjamin-Feir stability transition that $\mathscr{E}_2<\mathscr{E}_0<\mathscr{E}_1$ and therefore indicates that the downshifting is the most energetically viable state of the three. This grants a concrete explanation as to why downshifting may occur in the absence of viscosity: by downshifting, the Stokes wave is able to lower its wave energy and
 restabilise itself.

\subsection{Commentary on recurrence of the Stokes wave solution}
\label{subsec-recurrence}

In the majority of previous studies into the phenomena of frequency downshifting, it is argued that the shift in spectral peak to lower frequencies/wavenumber is a transient process and the system undergoes Fermi-Pasta-Ulam-Tsingou recurrence. This was primarily reported in \citet{yl82,b82} with the notion being refined in \citet{lm85} and \citet{hm91}, which was obtained by perturbing the carrier wave by the most (and only) unstable wave mode in the Nonlinear Schrodinger or Dysthe equation. However, once further sideband modes became unstable recurrence behaviour was lost completely and much more complicated dynamics occur (see the commentary of \S 6 of \citet{lm85} for their discussion on the matter). It transpires that this is also true if waves other than integer harmonics are initially excited within the system, where the wave-wave interactions cease to be closed, as in Zakharov equations \citep{j03,oos02}. We note that the heteroclinic connection we have proposed does not necessarily link a wave to its harmonic and thus we do no expect our interactions to be closed in the same way within our numerical procedures in \ref{sec:BR-sim}. This explains why, in the simulations within our paper regarding the presence of the heteroclinic wavenumber connection (\ref{R-front}), we do not observe any recurrence behaviour within the simulations.

On the other hand it is the case that the phase dynamics, in the neighbourhood of the
Benjamin-Feir transition, can capture recurrence behaviour. Primarily, oscillating solutions (corresponding to a back-and-forth transition between and initial and sideband wavenumber) can be obtained when the quartic potential associated with the travelling wave solutions of (\ref{BF-nondim}) has simple roots \citep{j09,kkl12}. As with the Gardner and mKdV equations, these can either be cnoidal or dnoidal solutions depending on which roots possess a valid connecting trajectory. It is more likely to be the former of these families responsible for the recurrence observed elsewhere, as the energy exchanges between modes are observed to have deeper troughs and sharper peaks; see, for example, Figure 3 in \cite{yf78} or Figure 13a in \cite{lm85}. 

The use of pseudospectral methods within this paper (and thus periodic boundary conditions) raises the question of numerical feasability of the phase dynamical solution outlined in \S \ref{sec:FD-sol}, as this would not be permissible in these numerical treatments as its boundary conditions would violate the periodicity required for spectral methods. This periodicity is not present experimentally, since this corresponds to an annular setup, and instead the energy is absorbed at the end of the tank and not redistributed within the wavetrain at later times. Thus, it is no surprise that pseudospectral numerical and Fourier-based approaches have thus far failed to explain conservative permanent frequency downshifts observed in experiments.  Moreover, the mechanism for dissipation, thought to be wavebreaking, is not observed until the wave steepness exceeds a certain threshold and the dissipative picture fails to adequately explain the presence of permanent downshifts in less steep waves.

An intermediary between these periodic solutions and the jump profile is the table-top solitary wave solution, arising when the potential possesses a double root and the remaining two roots are close to equal. In the context of phase dynamics, it represents a temporary shift in the wavenumber of a similar form to the jump solution discussed in \S \ref{sec:FD-sol} that eventually undergoes an inverse jump transition to the original wavenumber. Such solutions respect the periodicity requirement so long as the width of the tabletop solitary wave is less than the spatial domain. As such, one may repeat the travelling wave analysis for one repeated root and two that are $2 \delta$ apart:
\begin{equation}
    \bigg(\frac{dR}{d \xi^2}\bigg)^2 +\frac{\alpha_1}{2\alpha_3}(R-K^\infty)^2(R-K^0+\delta)(R-K^0-\delta) = 0\,.
\end{equation}
Here $K^\infty$ represents the far-field value of the solution and $K^0$ denotes the limiting value of the temporary wavenumber transition. The temporary downshifting solution corresponds to the case in which $K^0<K^\infty$, which we will restrict our discussion to. Comparisons with (\ref{BF-nondim}) give that these take one of the following pairs of values
\[
\begin{split}
K^\infty&= \frac{V}{2\alpha_1} \bigg(\alpha_2 + \sqrt{3\alpha_2^2-4\alpha_1-2 \bigg(\frac{\alpha_1\delta}{V}\bigg)^2}\bigg) = \bigg(-3.4721 + \sqrt{44.6262 - \frac{1}{2}\bigg(\frac{\delta}{V}\bigg)^2}\bigg)\, V\\[3mm]
K^0 &= \bigg(-3.4721 - \sqrt{44.6262 - \frac{1}{2}\bigg(\frac{\delta}{V}\bigg)^2}\bigg)\, V
\end{split}
\]
Thus, one may obtain the positive polarity solitary wave solution 
\begin{equation}\label{TT_wave}
R = 
K^\infty+\frac{(K^0-K^\infty)^2-\delta^2}{K^0 -K^\infty-\delta+\delta \cosh^2(1.0034\sqrt{(K^0-K^\infty)^2-\delta^2}\xi)}\,.
\end{equation}
As the parameter $\delta<0$ approaches zero, the solitary wave becomes the previously mentioned tabletop solitary wave with amplitude close to $K^0-K^\infty$.
This profile can be tested in numerical simulations in order to deduce its stability and robustness in the water wave problem via the use of reduced modelling.  Such discussion can be found in section \S \ref{sec:BR-sim} below.

\section{Numerical validation of the phase dynamics solutions}
\label{sec:BR-sim}

In order to verify the theoretical conclusions,
 arrived at from the phase dynamics analysis, we resort to a numerical investigation of a system representative of the water wave problem. We use the simplest water-wave model which contains the essential wave dynamics coupled to the meanflow, namely the Benney-Roskes system in (\ref{BR-1}-\ref{BR-3}).
The Benney-Roskes system possesses the same characteristic features, elliptic-hyperbolic transition, and phase dynamical picture as the full water wave problem in the case where $\mu,\,\tau$ in (\ref{Lag-WW}) are held fixed.
Towards this end, we expect the tabletop solitary wave solution (\ref{TT_wave}) to remain close to an exact solution within the Benney-Roskes system. 

To initialise the simulations, we use the solution (\ref{TT_wave}) in dimensional form and construct $A,B$ and $W$ according to (\ref{wave-quant-perts}) (accounting for the fact that the smallness of $a$ and $b$ has been factored out via $\epsilon$). This tabletop solution's width is chosen so that the local phase is periodic on a principle domain of length $L$ (typically, 80 wavelengths so that the tabletop is sufficiently flat), so that
\[
\int_{-\frac{L}{2}}^{\frac{L}{2}}U(\xi,T) \ d\xi = 0\,.
\]
We then inflate our computational domain by some factor of order 10, depending on the simulation time and distance from the Benjamin-Feir threshold, to ensure recursion does not occur in our numerical solution as discussed in \S\ref{subsec-recurrence}. Later wavenumber properties will then be computed on the principle domain of length $L$.
We choose $h_0\in [1.3, 1.3626]$ to ensure hyperbolicity of the underlying dynamics but to remain close to the modulation instability threshold and $\epsilon \sim 10^{-1}$ to align with typical experimental values for the steepness, and we note that this latter choice impacts simulation times due to its presence within the amplitude's evolution. We set $\eps \sim \epsilon^2$ to remain in the domain of asymptotic validity of the phase dynamics. For the time integration, we simulate in a periodic domain using an exponential time differencing scheme with Runge-Kutta timestepping of order 4 \citep{cm02} with the stability modifications outlined within \cite{kt05}. This numerical scheme has been verified a number of ways, including verification that the system conserves mass to $10^{-3}\%$ accuracy, that it admits the expected soliton solution when restricted to only the envelope equation (\ref{BR-1}) with $W=B=0$, and that when restricted to the shallow water components (\ref{BR-2}) and (\ref{BR-3}) with $A=0$ the solution generates two profiles which move at the characteristic speeds $ -c_g \pm \sqrt{g h_0}$. Momentum is not conserved in the Benney-Roskes system, as we discuss later within this section in relation to downshifting.

\begin{figure}
    \centering
    \begin{subfigure}{0.45\textwidth}
    \includegraphics[width=\textwidth]{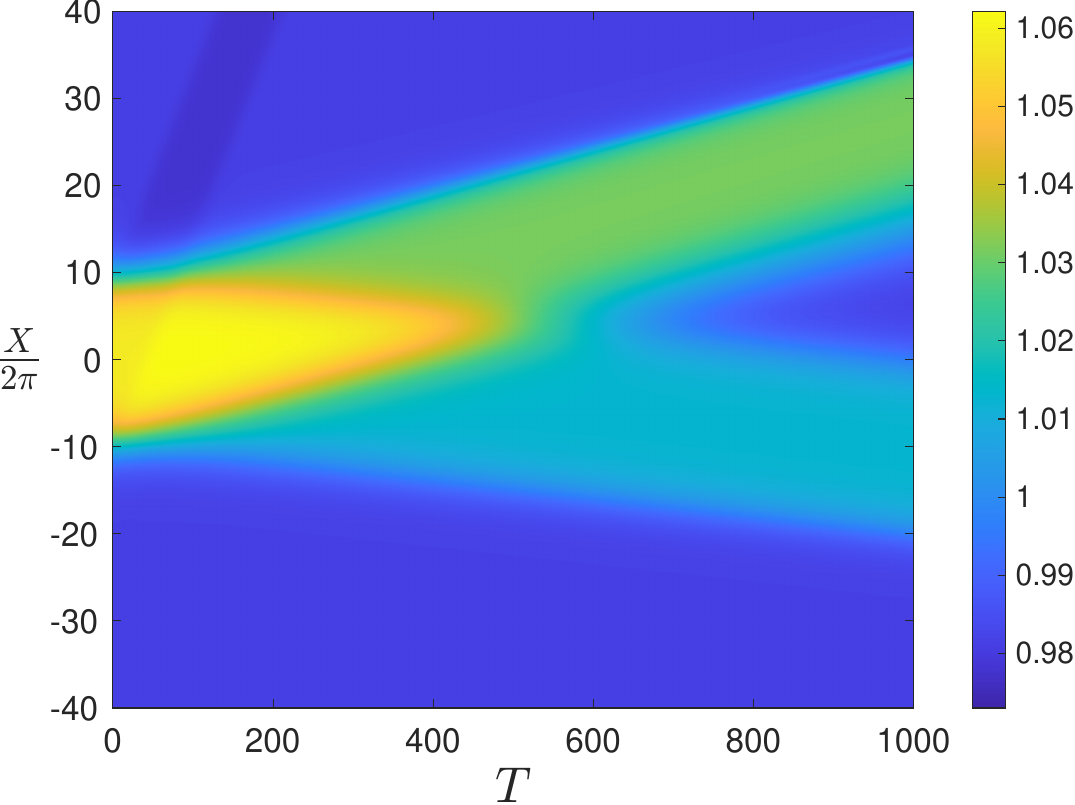}
    \end{subfigure}
        \begin{subfigure}{0.45\textwidth}
    \includegraphics[width=\textwidth]{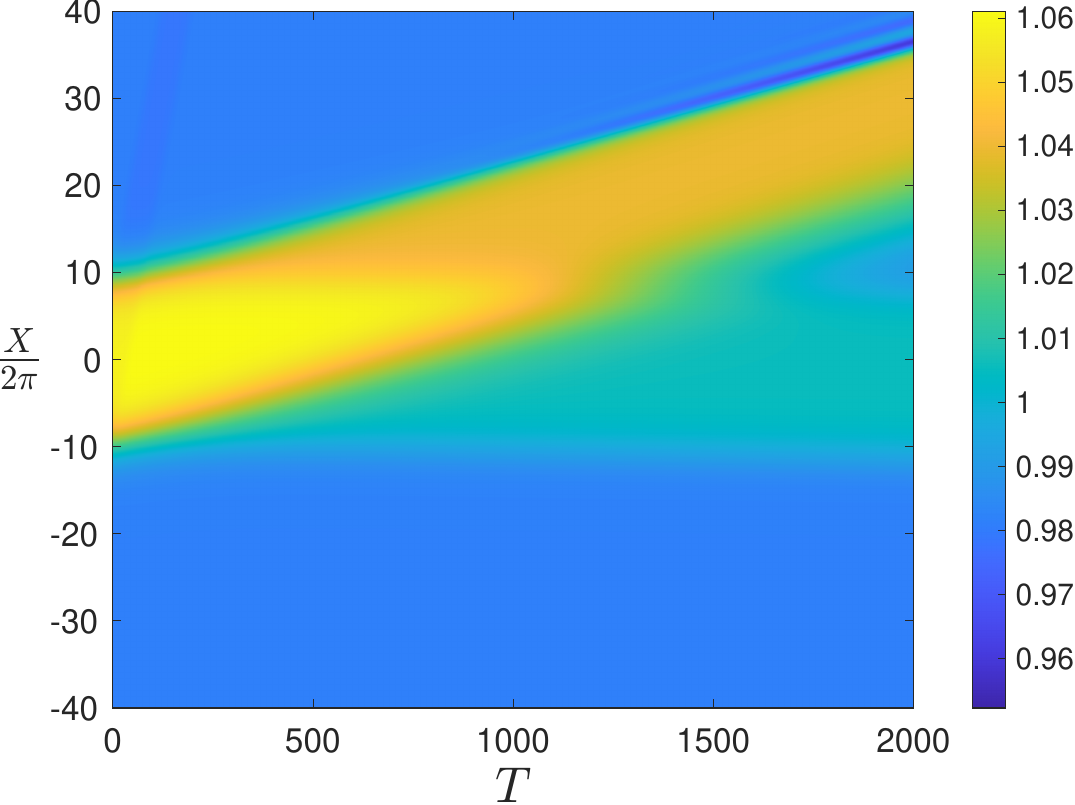}
    \end{subfigure}
    \caption{Space-time plots of the evolution of the wave envelope for $\eta_0 = 1.3$ (left) and $\eta_0=1.36$ (right). The initial tabletop splits into 4 components, each associated with one of the characteristics speeds of (\ref{BR-1}-\ref{BR-3})}
    \label{fig:amps}
\end{figure}

A representative example of the outcome of these simulations appears in Figure \ref{fig:amps}. What is observed is that the initial tabletop lump splits into four components, two associated with the long wavespeed of the shallow water component and two associated with the wave's group velocity. 
The latter two components emerge much later than the former as one would expect, and the amplitudes of these modes differ owing to the higher order terms present in the phase dynamical ansatz. All profiles maintain their general form within the simulation times, although one should note that the profiles associated with the group velocity can develop dispersive shocks at their leading edge if the initial data is large enough. Whilst this is likely indicative that the initial condition is outside the remit of the phase dynamic's validity, this shock formation does not significantly impact the observations related to the wavenumber discussed below. In fact, this appears to be more reflective of the experimental observations of downshifting observed by \citet{l77}, where the transition occurred via a modulation-demodulation cycle.

\begin{figure}
    \centering
    \begin{subfigure}{0.45\textwidth}
    \includegraphics[width=\textwidth]{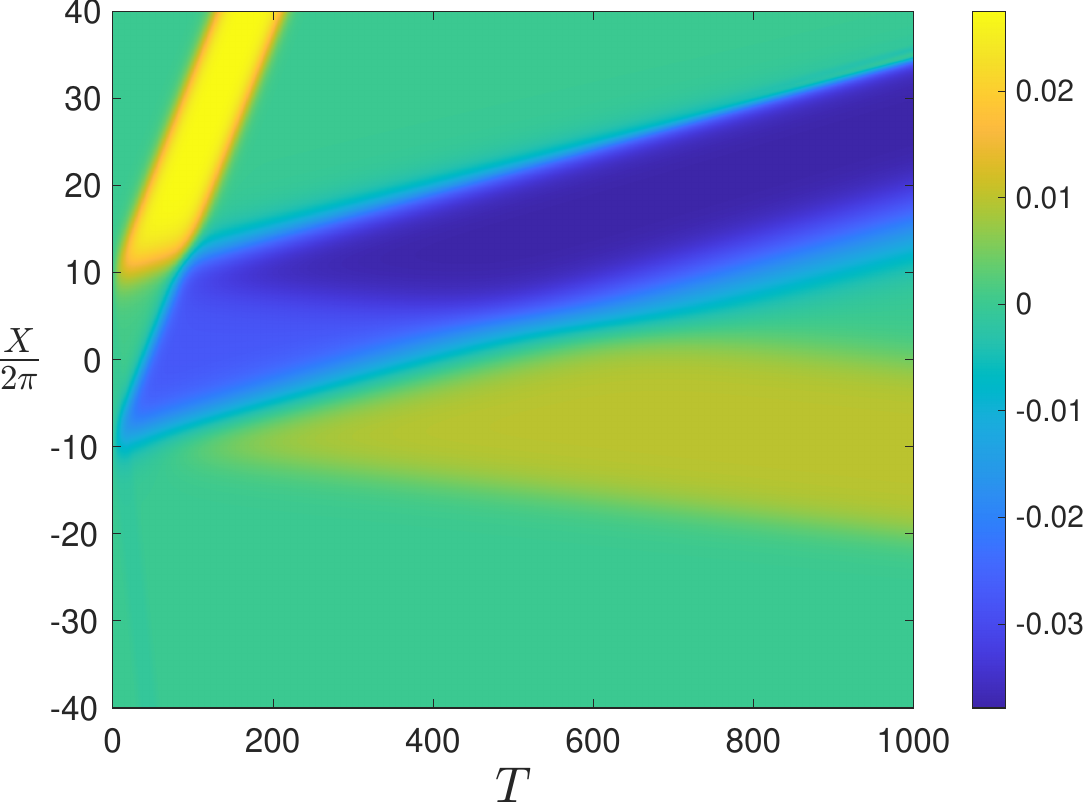}
    \end{subfigure}
        \begin{subfigure}{0.45\textwidth}
    \includegraphics[width=\textwidth]{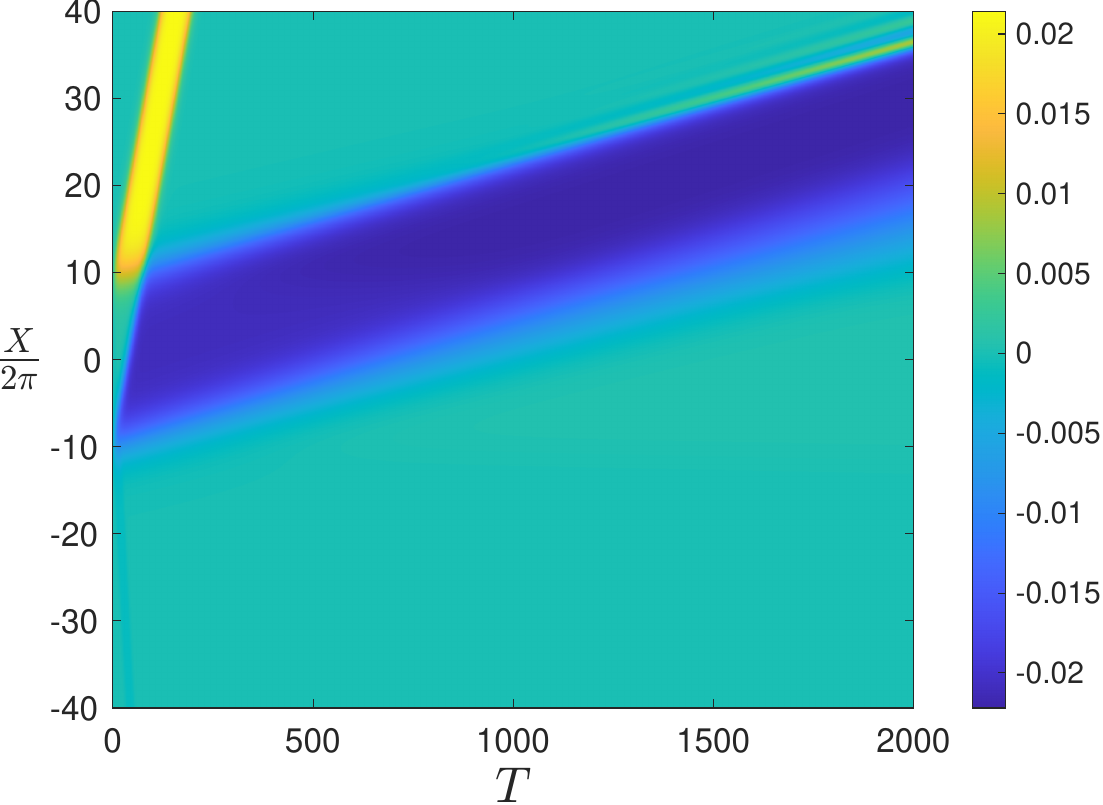}
    \end{subfigure}
    \caption{Local wavenumber associated with the profiles in Fig. \ref{fig:amps}, calculated via (\ref{local-wavenumber})}
    \label{fig:local-wavenumber}
\end{figure}

There are two key approaches we take to investigate wavenumber behaviour within the simulations. Our first approach is to extract the local wavenumber behaviour of the wave amplitude via the definition
\begin{equation}\label{local-wavenumber}
k_{local}(X,T) = {\rm Im} \left[\partial_X\ln \left(\frac{A}{|A|}\right)\right]\,,
\end{equation}
where ${\rm Im}$ indicates that the imaginary part is taken. An example of this extracted local wavenumber for a simulation appears in Figure \ref{fig:local-wavenumber}. We observe that there are three significant contributors to the local wavenumber change, two from the group velocity modes and one from the long wave mode. This latter mode emerges first and corresponds to an increase in wavenumber, and although this is not associated to the wavenumber transitions we intend to study it is not unexpected as this will correspond to another seemingly linear phase dynamic (cf. \S \ref{sec:PD}). Of the two modes associated
with the group velocity, the larger amplitude right-moving mode is responsible for the decrease in wavenumber and is significantly larger than the increase in local wavenumber of the other mode.

\begin{figure}
    \centering
    \begin{subfigure}{0.45\textwidth}
    \includegraphics[width=\textwidth]{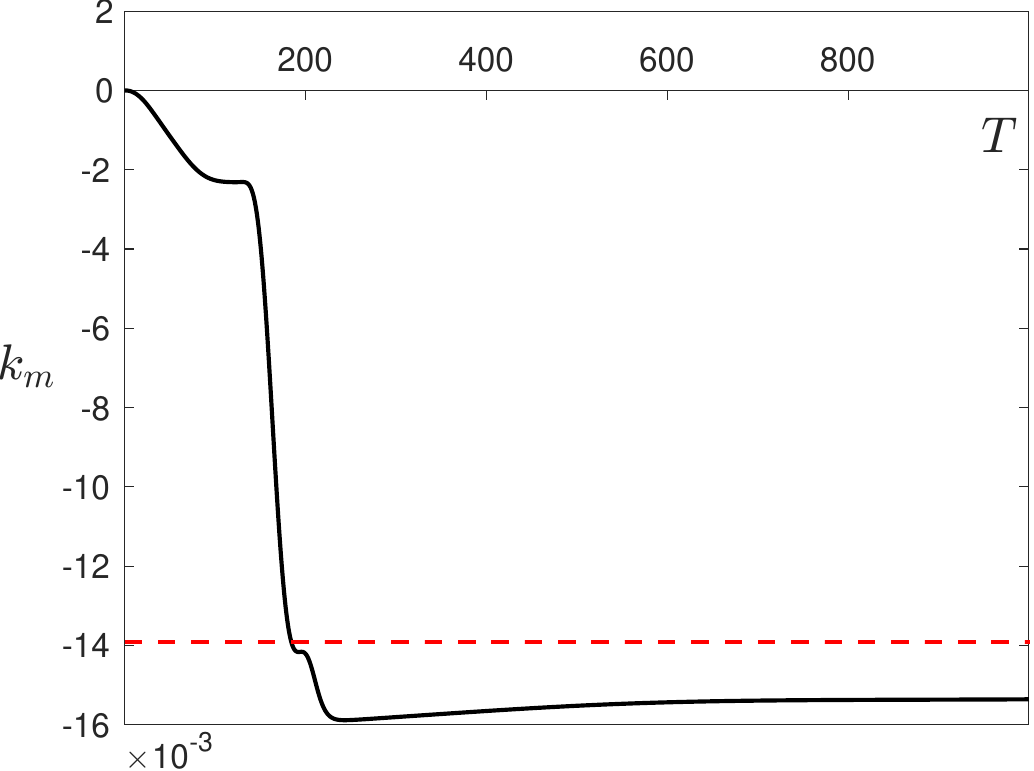}
    \end{subfigure}
    \quad
        \begin{subfigure}{0.45\textwidth}
    \includegraphics[width=\textwidth]{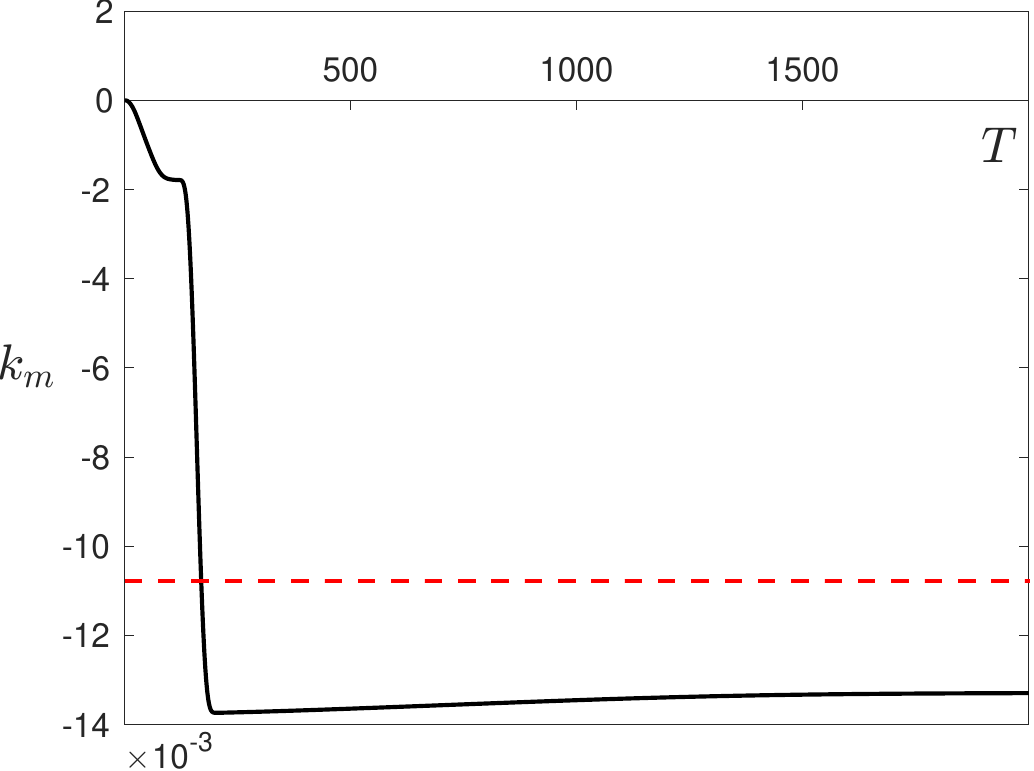}
    \end{subfigure}
    \caption{Spectral mean of the wavenumbers associated with the profiles in figure \ref{fig:amps} as a function of time. The dashed red line denotes the arithmetic mean of the long-time local wavenumber plateau values.}
    \label{fig:km}
\end{figure}

The second approach we take to determine wavenumber behaviour, particularly the emergent wavenumber behaviour for the entire wavetrain solution, will be to study the spectral mean, following \cite{ch19}.  It is defined by
\[
k_m(T) = \frac{i\int AA^*_X-A^*A_X \, dX }{2\int |A|^2 \, dX} \equiv \frac{\int k |\hat{A}|^2 \, dk}{\int |\hat{A}|^2 \, dk}
\]
where $\hat{A}(k,T)$ is the Fourier transform of the wave amplitude $A$. 
Whilst the momentum $\mathscr{P} = \int k|\hat{A}|^2 \, dk$ is conserved in the dynamics
of the nonlinear Schr\"odinger equation, it is not necessarily conserved under Benny-Roskes dynamics as
\[
\frac{d \mathscr{P}}{dT} = -2\left(k_0 W_X+\frac{gk_0^2-\omega_0^4}{2g\omega_0}B_X \right) |\hat{A}|^2
\]
We note the (total) wave mass (i.e. the denominator) is conserved under both the NLS dynamics and the Benney-Roskes dynamics, but the mass on the domain we examine the spectral mean decreases almost negligibly. This permits the spectral mean to change over the wavetrain's evolution, and we find that it decreases as depicted in Figure \ref{fig:km}. It depicts what is typical of a simulation with the prescibed setup: the wavenumber drops significantly as the long wave modes propagate out from the initial tabletop and subsequently out of the domain of interest, before slowly increasing to a negative asymptotic value. This value is reasonably close to the arithmetic mean of the local wavenumbers of each tabletop solution, suggesting that the spectral mean is the result of these wavenumber shifts. It confirms, however, that the tabletop solution is the source of the negative spectral mean value and that it remains stable over the course of it's propagation.

\begin{figure}
    \centering
    \begin{subfigure}{0.4\textwidth}
    \includegraphics[width=\textwidth]{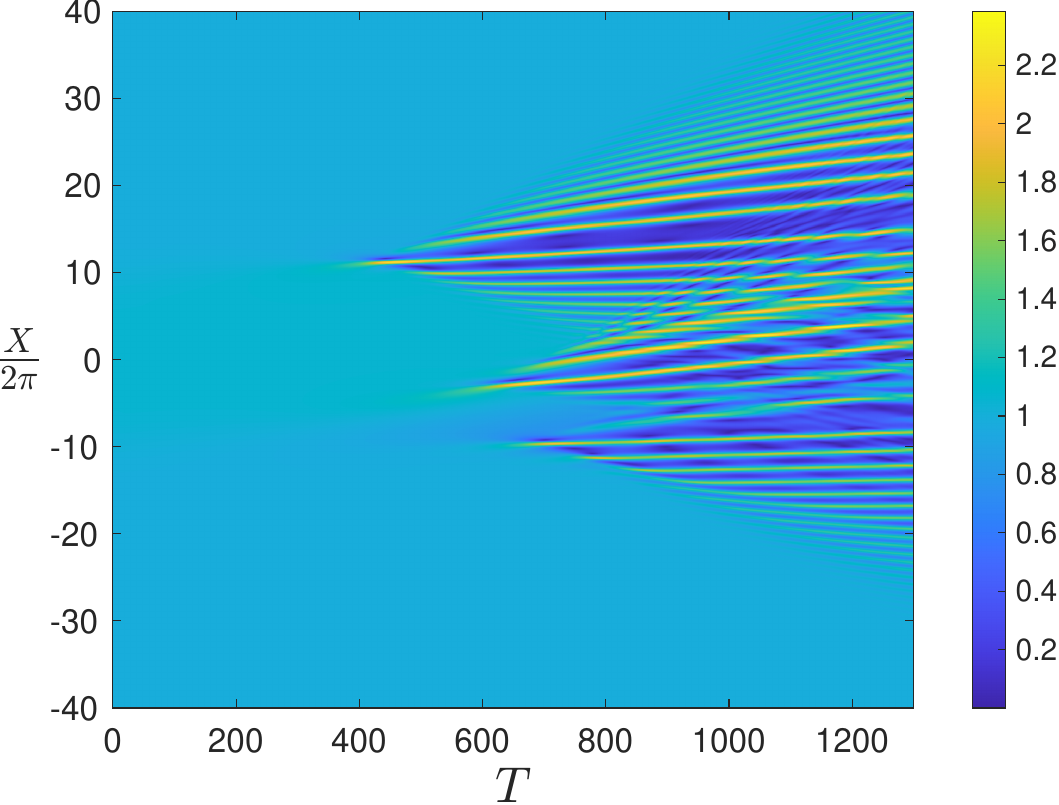}
    \end{subfigure}
    \quad
            \begin{subfigure}{0.4\textwidth}
    \includegraphics[width=\textwidth]{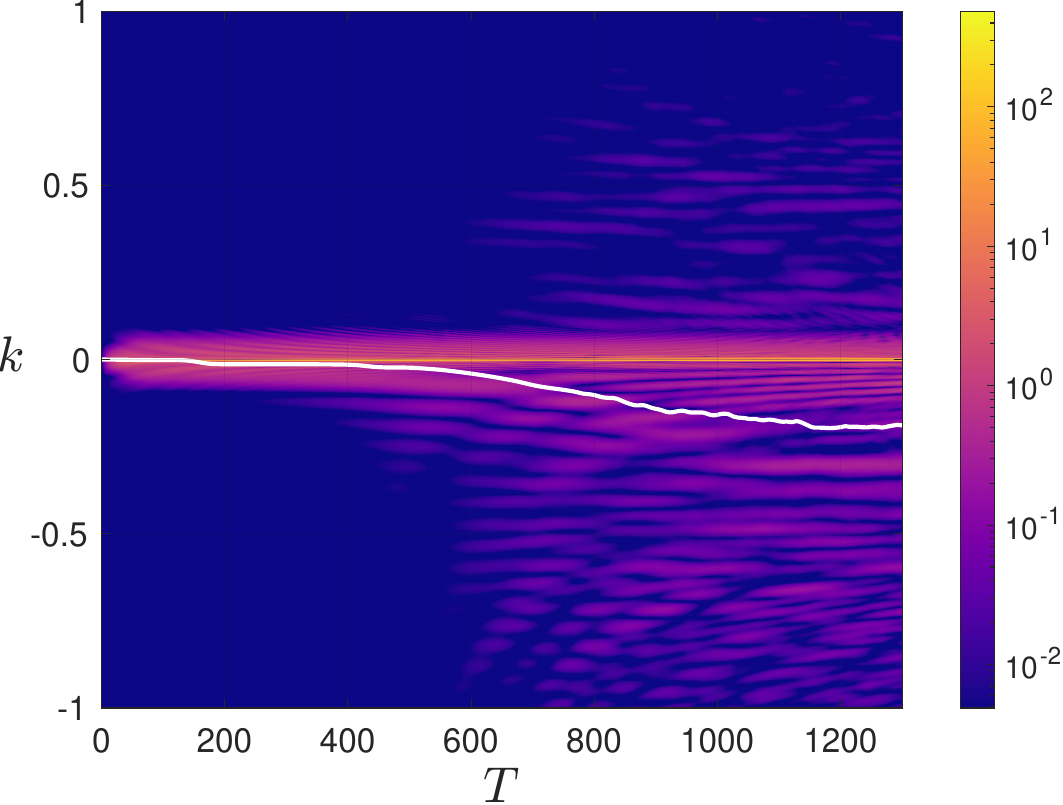}
    \end{subfigure}
            \begin{subfigure}{0.4\textwidth}
    \includegraphics[width=\textwidth]{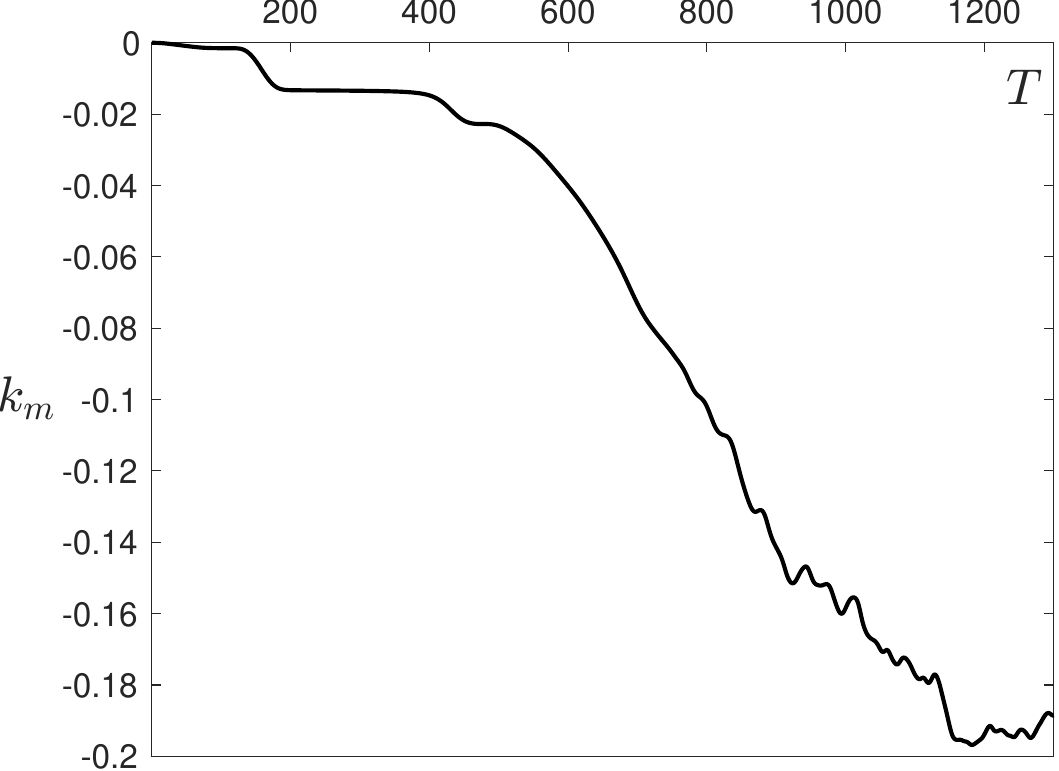}
    \end{subfigure}
    \caption{A numerical simulation for $\eta_0 = 1.4$, showing the amplitude (top left), power spectral density (top right) and , spectral mean wavenumber (bottom middle). The spectral mean wavenumber is marked with a white line on the power spectral density.}
    \label{fig:elliptic}
\end{figure}

What the Benney-Roskes system also allows us to do is to explore the elliptic (i.e. modulationally unstable) regime, unlike the phase dynamical description. What we do find, despite the phase dynamical description being invalid here, is that the same trend of frequency downshifting persists and in its initial stages follows the hyperbolic regime with an example appearing in Figure \ref{fig:elliptic}. As the modulation instability sets in, it works to improve the downshift substantially and decreases the wavenumber much further than the modulationally stable wavetrain close to threshold. We also observe that the transfer of energy in the power spectrum biases lower wavenumbers in line with the the observations from experiments \citep{l77,sb82,m82}. We reemphasise that there are no dissipative effects here - this bias towards lower wavenumbers in the numerical simulation is entire mean-flow driven. It does however remain an open question whether the wave profiles within these simulations become steep enough to break and thus for dissipation to play a role in this regime, but these simulations show that this is not required for the downshifting phenomena in the modulationally unstable regime. What we do not see in this case is recurrence or restabilisation, instead seeing the formation of a soliton train which initially forms at the edges of the splitting tabletop solitary wave and expands over the simulation time. 

\section{Concluding Remarks}
\label{sec-cr}

This paper has introduced a mechanism for water waves to undergo a permanent frequency downshift without dissipative effects. Mathematically, it is due to a local loss of genuine nonlinearity arising at the Benjamin-Feir transition that introduces higher order effects that support front profiles for the wavenumber's evolution. Energetically, it occurs because of a decrease in energy which restabilises the Stokes wave. This energetic perspective helps to reinforce the observations that the energy exchange to the lower sideband is much higher than that to upper sidebands. This paper highlights that these effects are due to mean flow effects present in the problem.

We re-emphasise that, whilst the explanation for downshifting without the reliance on dissipation is an important step forward in our understanding of the phenomenon, dissipation remains one possible (and provably successful) mechanism for  decreases in spectral peaks~\citep{cg16,ch19}. Particularly, the dissipative description is valid deep into the elliptic regime, well outside of the transition regime which the phase dynamical description of this paper is applicable. An important future avenue for study will be to compare the interplay of these effects and in which regimes one effect may dominate over the other and to the degree in which it does so.

The assumption of gravity waves is important here, as it ensures that $\omega''_0 \neq 0$ for any choice of $k$ and thus ensures the nondegeneracy of the phase dynamics. The
inclusion of capillary effects introduces a new modulation stability boundary at points where $\omega_0'' = 0$, and such points cause every coefficient in (\ref{kdv-SW}) to become zero. To rebalance the phase dynamics in this case, one must include higher order dispersive effects in addition to the considerations made for the scenario of this paper. The result is a fully extended version of the two-way Boussinesq which supports localised solutions as well as fronts. 

The mechanism proposed here is universal in that it does not rely on
a particular governing equation, 
just on the fact that the equations are generated
by a Lagrangian and there exists a multi-parameter family of periodic
or multi-periodic travelling waves, with parameter values at which
two characteristics coalesce, and furthermore the nonlinearity in the
re-modulation is cubic.  Indeed, generalisations of the water wave problem
(inclusion of surface tension, variable density, electromagnetic fields, etc),
at both finite amplitude and weakly nonlinear limit,
would be settings where the above scenario is likely to occur.

Although downshifting has been shown to be the only energetically viable result in the gravity water wave problem, upshifts
are also theoretically possible in other physical systems. 
For example, Whistler waves in the magnetosphere may undergo a downshift or an upshift depending on whether they are travelling parallel or orthogonal to the background magnetic field~\citep{oks08,o21,ra23}. The analysis performed here is likely to explain when this is permissible and will be due to how the amplitude varies with the mean flow element (which for plasmas takes the role of velocity and number density perturbations), in comparison to the principal change in the wavenumber and frequency.

\subsection*{Acknowledgements}
D.J.R would like to thank the participants of the \emph{Dispersive Hydrodynamics}, especially Pat Sprenger and Paul Milewski, for their invaluable discussions throughout the development of this work.

\subsection*{Funding}
The authors would like to thank the Isaac Newton Institute for Mathematical Sciences for support and hospitality during the programme \emph{Dispersive Hydrodynamics} when work on this paper was undertaken. This work was supported by: EPSRC grant number EP/R014604/1

\subsection*{Declaration of Interest}
The authors report no conflict of interest.

\appendix

\section{Effect of mean velocity on the amplitudes $a$ and $b$}
\label{app-ab}

The perturbative impact on the amplitude and mean flow of the Stokes wave,
in the neighbourhood of fixed values $(a_0,b_0)$, is given in this appendix.
Using the definitions (\ref{RE-defn}), (\ref{char-ww}) and (\ref{eigenvect}), we can show that the leading order contributions to their change is
\begin{equation}\label{LO-perts}
\begin{gathered}
a = a_0+\frac{gh_0+B_0\omega_0'}{c_0h_0\Delta_W}\eps \zeta_2U+\mathcal{O}(\eps \sqrt{E},\eps^2)\,, \\ b = b_0-\frac{1}{\omega_0\Delta_W}\bigg(\omega_2^0\omega_0'+\frac{B_0k}{c_0h_0}\bigg)\eps \zeta_2 U+\mathcal{O}(\eps \sqrt{E},\eps^2)\,,
\end{gathered}
\end{equation}
where $\zeta_2$ is the second component of the eigenvector $\be$ in (\ref{E-def}). 
This expansion
highlights that the primary effect on the Stokes waves arises from the mean flow element of the problem, evident from the appearance of $\zeta_2$, as opposed to being driven by the wave itself.  The expressions preceding the perturbation $U$ represent recurring factors within the analysis, and so it will be convenient to define the two quantities
\begin{equation}\label{M-defns}
\mathscr{M}_1 = \frac{gh_0+B_0\omega_0'}{c_0h_0\Delta_W}<0\,, \qquad \mathscr{M}_2 = -\frac{1}{\omega_0\Delta_W}\bigg(\omega_2^0\omega_0'+\frac{B_0k}{c_0h_0}\bigg)>0\,.
\end{equation}
These quantities arise due to the variations of the wave amplitude and bulk flow due to the mean-flow effects and will be important in characterising the impact of these effects on the dynamics of the wave.
Evaluating these perturbations to the wave amplitude and bulk flow at the Benjamin-Feir transition one finds
\begin{equation}\label{wave-quant-perts}
a = a_0-0.9171 \eps U+\mathcal{O}(\eps \sqrt{E},\eps^2)\,, \qquad b = b_0+0.8220 \eps U+\mathcal{O}(\eps \sqrt{E},\eps^2)
\end{equation}
As $\zeta_1>0$ at the transition, one can infer that increases in the wavenumber, corresponding to positive $U$, result in a decrease of amplitude and a rise (drop) in mean level, and vice versa. This is in line with the experimental observations of \cite{l77}. We may also use this information to infer the stability of the Stokes wave by assessing how the nondimensional depth $kh$ changes near this transition. To leading order, this is
\[
kh = kh_0+2.0829\eps U+\mathcal{O}(\eps \sqrt{E},b,\eps^2)\,.
\]
As expected, the states which increase the wavenumber also increase $kh$, but do so at a faster rate than the changes in the mean level that would otherwise balance it out.

\section{Explicit expressions for matrix pencil entries}
\label{app-E-Expr}

Explicit expressions for the entries of the matrix pencil ${\bf E}(c)$ defined in (\ref{E-def}) are given here,
\begin{equation}
\begin{split}
E_{11} &= \frac{g(c-c_g)^2}{\omega_0^2\Delta_{W}} +\bigg[\frac{\omega_0''}{\omega_0}+\frac{c-c_g}{\omega_0}\bigg(\frac{(c-c_g-2\omega_0')}{\omega_0}+\frac{2}{c_0h_0\Delta_W}\big(B_0\mu'-g\tau '\big)\bigg)\bigg]E\\
&\hspace{5cm}-\frac{2g(c-c_g)}{\omega_0\Delta_W}\mu'kb\,,\\[3mm]
E_{12} &=-\frac{(c-c_g)(g h_0+B_0 (c-u))}{c_0^2kh_0\Delta_{W}}\\
&+\bigg[\frac{c-c_g-\omega_0'+c_0}{c_0^2 k}+\frac{(gh_0+B_0(c-u) )}{c_0h_0\Delta_W}\tau'-\frac{1}{\Delta_W}\bigg(\frac{ kB_0}{c_0h_0}+\omega_2^0 (c-u) \bigg)\mu'\bigg]E\\
&\hspace{5cm} +\frac{(gh_0+B_0(c-u) )}{c_0h_0\Delta_W}\mu' b\\[3mm]
E_{22} &=\frac{gh_0+2B_0(c-u)+B_0^2}{c_0^2h_0\Delta_{W}}-\frac{\omega_2(gh_0-(c-u)^2)}{\omega_0\Delta_{W}}+b-\frac{E}{c_0^2}\,.
\end{split}
\end{equation}
This matrix has a right eigenvector associated with the characteristic (\ref{char-ww}) defined by ${\bf E}(c)\be=0$.  It can be
expanded for small amplitude to give
\begin{equation}\label{eigenvect}
\begin{split}
\be = \begin{pmatrix}
\zeta_1\\
\zeta_2
\end{pmatrix} =& \pm \sqrt{\omega_0'' \omega_2^{eff}}
{\bm \chi}\\
&+ \sqrt{E}\left[C_2{\bm \chi}-\begin{pmatrix}
\frac{B_0 \omega_0'' \omega_2^{eff}}{c_0h_0\Delta_W}+\frac{\omega_0'}{c_0}-1+\frac{1}{\Delta_W}\big(\omega_2^0\omega_0'+\frac{kB_0}{c_0h_0}\big)\mu'-\frac{k(B_0\omega_0' + gh_0)}{h_0\Delta_W}\tau'\\
 \omega_0''\big(1+\frac{g \omega_2^{eff}}{\omega_0\Delta_W}\big)
\end{pmatrix} \right]\\[5mm]
&\pm E \left[C_3 {\bm \chi} +\sqrt{\omega_0'' \omega_2}\left( 
\begin{pmatrix}
\frac{1}{c_0}\\
0
\end{pmatrix}+\frac{2C_2}{\Delta_W}
\begin{pmatrix}
 \mu\\
-\frac{g}{\omega_0} 
\end{pmatrix}
\right)\right]\\
&+ \mathcal{O}(E^{\frac{3}{2}})\,,
\end{split}
\end{equation}
with
\[
{\bm \chi} = -\frac{1}{c_0h_0\Delta_W} \begin{pmatrix}
gh_0+B_0\omega_0'\\ 0 
\end{pmatrix}\,,
\]
where $C_3$ is a further term in the amplitude expansion of the characteristic.
It is not given here as it does not contribute to the analysis at the orders considered.

\bibliographystyle{jfm}
\bibliography{Arxiv_Ver.bib}


\end{document}